\documentclass[letterpaper,12pt]{article}
\usepackage{osajnl2}
\usepackage[draft]{hyperref} 
\usepackage{url}

\begin{document}


\title{Effects of thermal gradients on total internal reflection corner cubes}
\author{S.\,D. Goodrow$^{1}$ and T.\,W. Murphy, Jr.$^{1,*}$}
\address{$^{1}$Center for Astrophysics and Space Sciences, University of California, 
San Diego, \\ 9500 Gilman Drive, MC-0424, La Jolla, CA 92093-0424, USA}
\address{$^{*}$Corresponding author: \tt{tmurphy@physics.ucsd.edu}}


\begin{abstract}

Uncoated corner cube retroreflectors (CCRs) operating via total internal reflection (TIR) are less susceptible to internal heating than their metal-coated analogs, lacking an absorber on the rear surface.  Even so, environments that induce differential heating within the CCR will result in thermal lensing of the incident wavefront, introducing aberrations that will generally reduce the central irradiance of the polarization-sensitive far-field diffraction pattern (FFDP). In this paper, we characterize the sensitivity of TIR CCRs to axial and radial thermal gradients.  We present simulated FFDPs for key input polarizations and incidence angles and provide a generalized analytic model that approximates the behavior of the central irradiance as temperature differences are introduced.

\end{abstract}


\ocis{220.1010, 260.1960, 260.2710, 260.5430, 260.6970, 350.6830}
\maketitle


\section{Introduction}

Total internal reflection (TIR) corner cube retroreflectors (CCRs) are
often advantageous choices in the presence of significant radiative
flux---such as sunlight---due to their lack of absorptive coatings. Still,
these optical devices can become subject to thermal gradients in the
presence of dust deposition, surface abrasions, bulk absorption of light
within the material, and conduction from their mounting arrangement. Such
differential heating will produce distortions of the emerging wavefront as
a form of thermal lensing on their far-field diffraction pattern (FFDP). It
is important then to understand the sensitivity of the FFDPs produced by
TIR CCRs to perturbations from thermal gradients. Our specific interest
arises from observations of the degraded performance of the retroreflector
arrays placed on the Moon by the Apollo astronauts \cite{dust}.  Although
we concentrate on the impact of thermal lensing on TIR CCRs, many of our
results apply to the metal-coated variety as well.

The literature contains a number of papers that mention thermal sensitivity
of TIR CCRs, however none provide a full quantitative analysis. A
contribution to the final report on the Apollo 11 Laser Ranging
Retro-reflector Experiment by Emslie and Strong (1971) \cite{nasa} provides
the closest quantitative handle, however lacks the computational power to
give a complete view of how the FFDPs evolve as a function of thermal
gradient and angle of incidence. Faller (1972) \cite{faller} reports on the
optimal dimensions for CCRs in retroreflector arrays, including thermal
considerations, but does not detail their subsequent thermal lensing
behaviors.  Zurasky (1976) \cite{zurasky} presents experimental findings
for combined axial and thermal gradients in CCRs employing dihedral angles,
and evaluating flux in an annulus well away from the central spot.  More
recently, in efforts to produce next-generation retroreflector arrays,
Currie et al.  (2011) \cite{currie} mention the importance of minimizing
thermal gradients to improve return signals. Dell'Agnello et al. (2011)
\cite{dellagnello} further emphasize the degradation of the return signal
from thermal lensing, but neither provide generalized results on the
subject.

The goal of this paper is to provide a general characterization of the
FFDPs produced by TIR CCRs subjected to axial and radial thermal gradients,
with application to our lunar ranging project \cite{apollo}.  The
simulations herein build directly on our analysis framework for evaluating
TIR CCRs (companion paper \cite{raytrace}), and we make the computer code available online.

The quantitative results here target thermal conditions relevant to space
environments, while providing a framework that can be applied to 
other conditions. Additionally, we display thermally-lensed FFDPs for a
variety of key input polarizations and incidence angles. We detail the
methodology and include analytic solutions to approximate the behavior of
the central irradiance as thermal gradients are introduced at normal
incidence.


\section{Method}\label{sec:method}

While all paths through an isothermal CCR for a given angle of incidence
have the same geometric pathlength, $l$, a solid glass CCR subject to
thermal gradients will experience thermal expansion leading to
path-specific deviations, $l^{'} = l \left( 1 + \alpha \overline{\delta T}
\right)$, where $\alpha$ is the thermal expansion coefficient, $\delta T$ is the departure from some reference temperature, $T_0$, and the bar represents a path average.
Additionally, the average wavelength within the medium,
$\overline{\lambda}$, can differ for each path as the refractive index
varies throughout the CCR according to the thermo-optic
coefficient, $\beta$:
\begin{equation}
\overline{\lambda} = \frac{\lambda_{0}}{n_{0} + \beta \overline{\delta T}},
\label{eq:average-wavelength}\end{equation}
where $\lambda_0$ is the vacuum wavelength and $n_0$ is the refractive index at temperature $T_0$.

Consequently, the emerging wavefront will have nonuniform phase shifts
leading to thermal lensing of the FFDP. The absolute phase in radians can
be calculated for an arbitrary path from its average wavelength within the
medium:

\begin{equation}
\phi = \frac{2 \pi l^{'}}{\overline{\lambda}} = \frac{2 \pi \, l \left( 1 + \alpha \overline{\delta T} \right) \left(n_{0} + \beta \overline{\delta T}\right)}{\lambda_{0}}
\label{eq:phase-distortion}\end{equation}

Expanding this and neglecting second-order terms in $\delta T$ allows us to define a
generalized thermal coefficient $\eta \equiv \alpha n_{0} + \beta$, and rewrite
Equation~\ref{eq:phase-distortion} as

\begin{equation}
\phi = \frac{2 \pi l}{\lambda_{0}}\left( n_{0} + \eta \overline{\delta T} \right).
\label{eq:phase-distortion-final}\end{equation}

As our analysis is aimed at the use of retroreflectors in the lunar daylight environment, we are particularly interested in the sensitivity of fused silica CCRs to thermal gradients at temperatures in the range of 300--350~K, which have a refractive index of $n_{0} = 1.46071$ at our chosen wavelength of $\lambda = 532$~nm. We therefore use $\alpha \approx 5 \times 10^{-7}$~K$^{-1}$ and $\beta \approx 10^{-5}$~K$^{-1}$ in our analysis \cite{dndt, nasa}. We see that thermal expansion plays a minor role in thermal lensing for Apollo CCRs and use $\eta = 10^{-5}$~K$^{-1}$ in our analysis to provide easy scaling of our results to other conditions.

The approximate cylindrical symmetry of the circularly cut CCR suggests two
primary modes of thermal gradient distribution: axial and radial. As such,
we modeled axial temperature differences, $\Delta T_{h}$, from the CCR's
front face to its vertex at the origin, radial temperature differences,
$\Delta T_{R}$, from the CCR's center to its periphery, as well as
superpositions of the two. We use the geometry of the Apollo CCRs,
$R=19.05$~mm and $h=29.80$~mm, in our simulations and numerical results. As
the radial displacement, $r = \sqrt{x^{2} + y^{2}}$, is independent of $z$,
the two gradients are orthogonal and can be computed separately given the
pathlength-weighted averages $\overline{z}$ and $\overline{r}$ of the ray
path in question. The average temperature offset along the path is then:
 
\begin{equation}
\overline{\delta T} = \frac{\Delta T_{h}}{h} \overline{z} + \frac{\Delta T_{R}}{R} \overline{r}
\label{eq:thermal-model}
\end{equation}

Both $\overline{r}$ and $\overline{z}$ have analytic expressions given the coordinates of the endpoints of each path segment. We can determine the approximate peak-to-valley phase difference of the wavefronts by calculating the difference of the phase offsets for normally-incident paths through the center (subscript c) and perimeter (subscript p) of the CCR, under the assumption that the thermal lensing for both axial and radial thermal gradients will be dominated by the difference between central and peripheral paths. If the absolute difference in mean temperature between these two paths is $\Delta \overline{\delta T} = \left| \overline{\delta T}_{\mathrm{c}} - \overline{\delta T}_{\mathrm{p}} \right|$, then the phase offset difference is

\begin{equation}
\Delta \phi = \frac{2 \pi l \eta \Delta \overline{\delta T}}{\lambda_{0}}.
\label{eq:phase-dif}
\end{equation}

The central path has two path segments of equal length, from the front face
to the vertex and back, making $\overline{z_{\mathrm{c}}} = \frac{h}{2} \approx
14.90 \mathrm{\;mm}$ and $\overline{r_{\mathrm{c}}} = 0 \mathrm{\;mm}$. The
perimeter path is dependent on azimuth of entry, so we choose the path that
enters above an edge of the CCR. This path has three path segments of
unequal length, leading to pathlength-weighted averages $\overline{z_{\mathrm{p}}}
\approx 8.83$~mm and $\overline{r_{\mathrm{p}}} \approx 12.59$~mm. Thus the
perimeter ray stays closer to the front surface and obviously spends time
farther from the center than does the central ray. If we arbitrarily set
$\Delta T_{h} = \Delta T_{R} = 1$~K,
the differences
of the average temperature offsets of the two paths are $\Delta
\overline{\delta T}_{\mathrm{axial}} = 0.20$~K and $\Delta \overline{\delta
T}_{\mathrm{radial}} = 0.66$~K. This yields peak-to-valley phase
differences of $\Delta \phi_{\mathrm{axial}} = 1.43$~radians and $\Delta
\phi_{\mathrm{radial}} = 4.66$~radians. In the radial gradient case, the
wavefront perturbations vary with azimuth and our result corresponds to a
minimum. The maximum, $\Delta \phi_{\mathrm{radial}} = 5.11$~radians,
occurs for paths that enter or exit at an azimuth $30^{\circ}$ away from an
edge.

\begin{figure}
\includegraphics[scale=0.4]{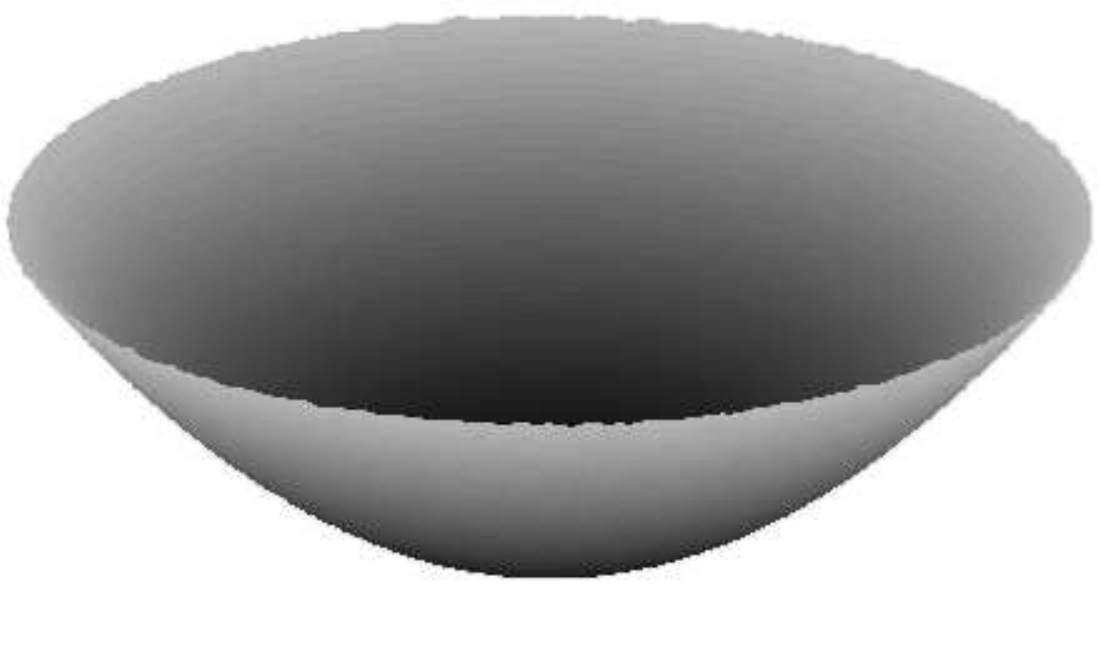}\hfill{}
\includegraphics[scale=0.4]{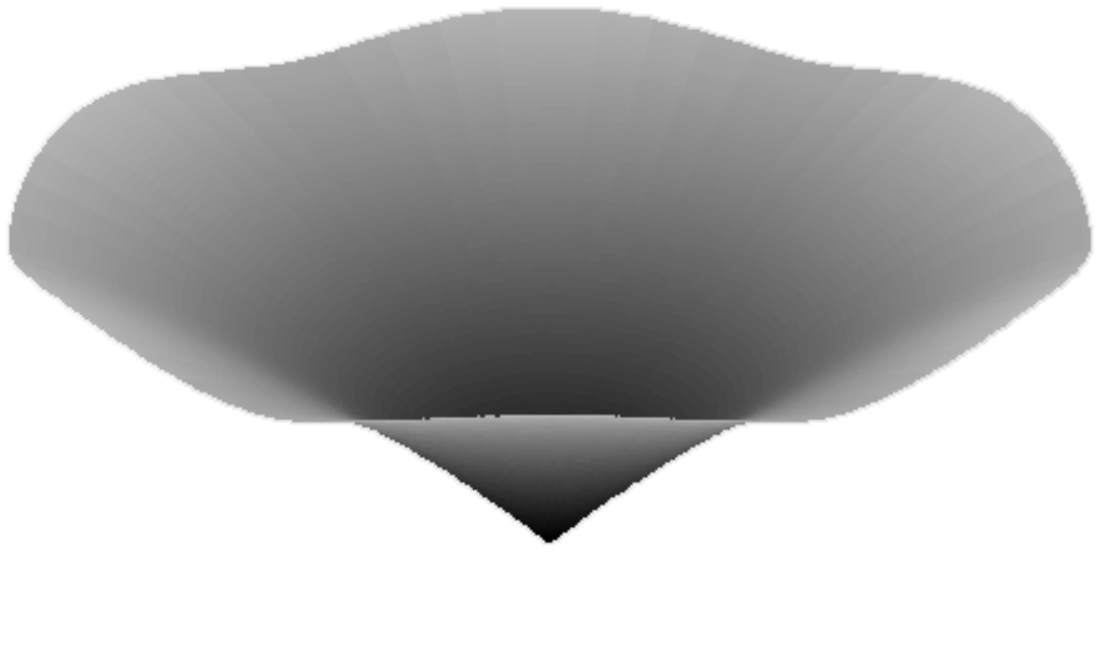}\hfill{}
\includegraphics[scale=0.4]{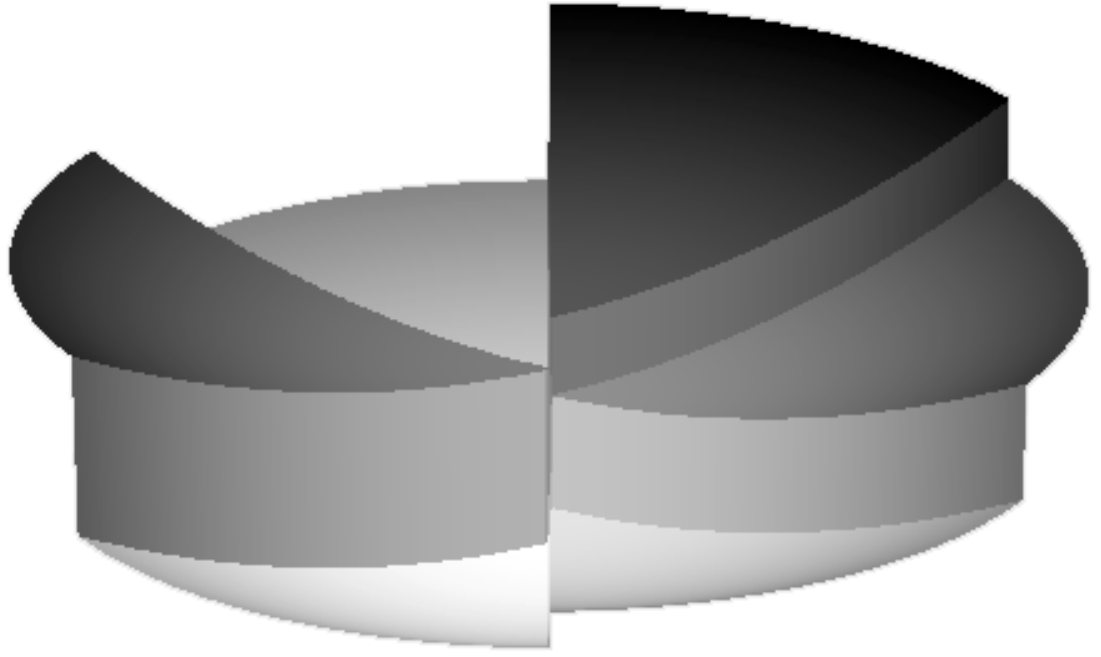}

\caption{Three-dimensional representations of wavefronts produced by a CCR
subject to a thermal gradient at normal incidence. At left is the perfect
reflecting (metal-coated) case for an axial thermal gradient, resulting in a perfectly
spherical wavefront. At center is the perfect reflector afflicted with a
radial thermal gradient. At right is an example TIR case, depicting the
combined effects of the TIR phase shifts and the perturbations from an
axial thermal gradient.  If the wavefront travels upward, all panels
correspond to the corner cube being cooler on the front surface or on the
periphery---although the results of this paper are insensitive to the sign
of the thermal gradient.} \label{fig:3d-wavefronts}

\end{figure}

We employed a ray tracing algorithm in Python to track the phase offsets of
each path through the CCR \cite{raytrace}. In the case of a metal-coated 
CCR---assumed to have
perfect reflection at its rear surfaces---the phase of the input wave is
shifted uniformly by $\pi$ radians at each interface. Consequently, the
departure of the output wavefront from planarity deriving from a planar input wave can be used to
characterize the effects of the thermal gradient on the wavefront. Such is
not the case for TIR CCRs, as TIR at the rear surfaces induces phase shifts
dependent on the angle of incidence. Here, each of the six wedges within
the aperture acquires a potentially different phase shift in addition to
the smooth perturbations from the thermal gradient. Three-dimensional
representations of these wavefronts are shown in
Figure~\ref{fig:3d-wavefronts} for normal incidence.

\section{Simulation Analysis}
The orientation of our FFDPs represent direction cosines in the global frame,
as specified in our companion work \cite{raytrace}. In this frame, the CCR is oriented to have a real edge extending toward the right. Horizontal and vertical polarization states are likewise defined according to the same reference.

\begin{figure}
\includegraphics[scale=0.85]{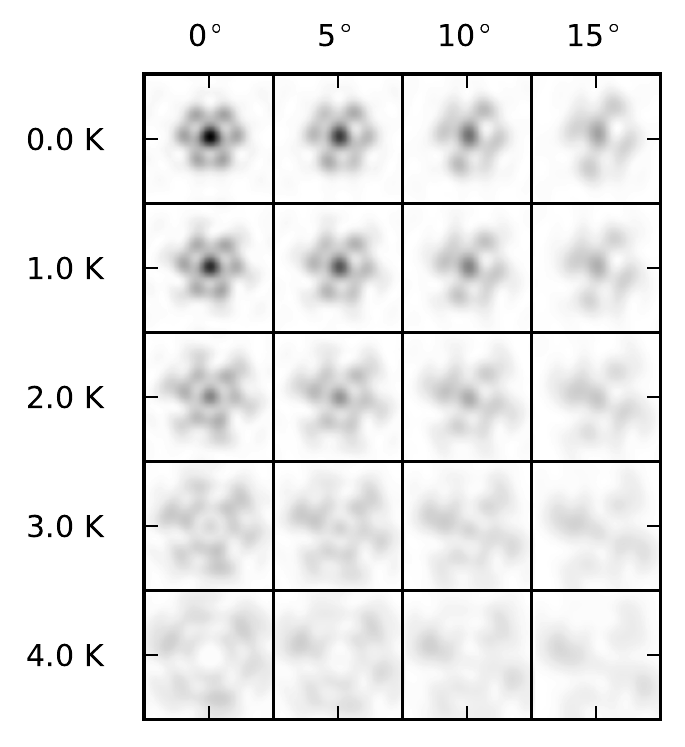}\hfill{}
\includegraphics[scale=0.85]{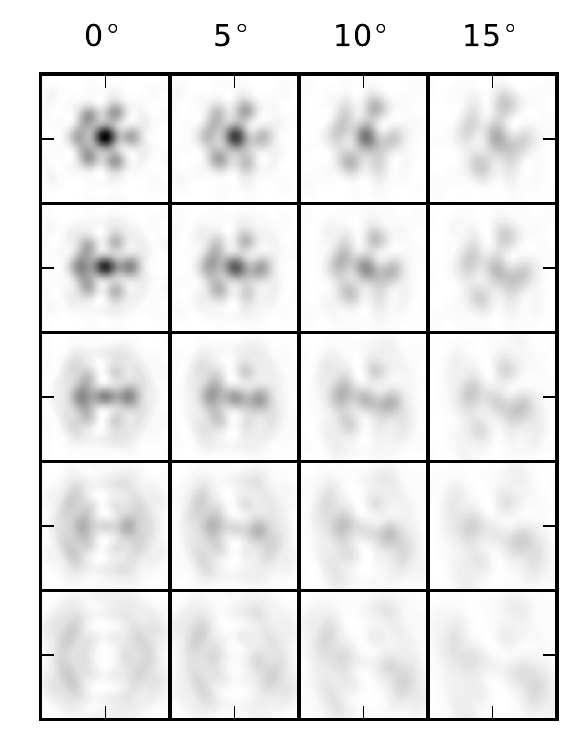}\hfill{}
\includegraphics[scale=0.85]{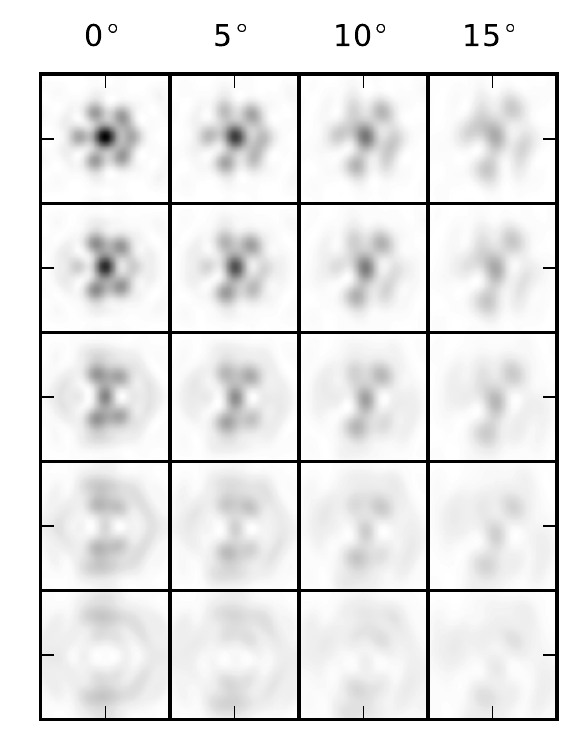}
\caption{FFDPs produced by a TIR CCR with an axial thermal gradient. Right-handed circular input polarization is at left, followed by horizontal and vertical linear input polarizations. In each grid, the incidence angle increases in $5^{\circ}$ steps from left to right.  Rows correspond to various front-to-vertex temperature differences, increasing in 1.0~K steps. Each frame is $50\lambda/7D$ radians across. Irradiance is normalized to the normal incidence isothermal frame in the upper-left.  All results herein correspond to the behavior of fused silica at a wavelength of 532~nm.}
\label{fig:ffdp-axial-grid}
\end{figure}

\begin{figure}
\includegraphics[scale=0.85]{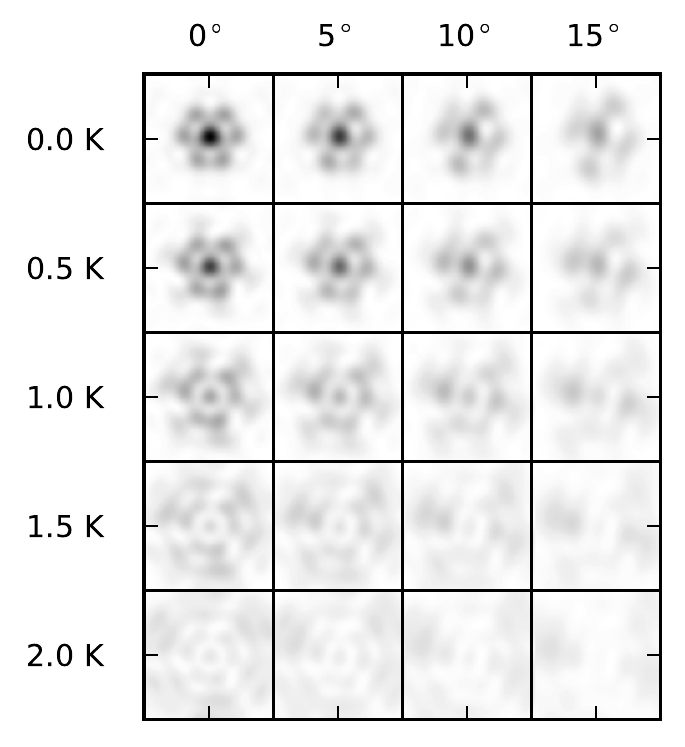}\hfill{}
\includegraphics[scale=0.85]{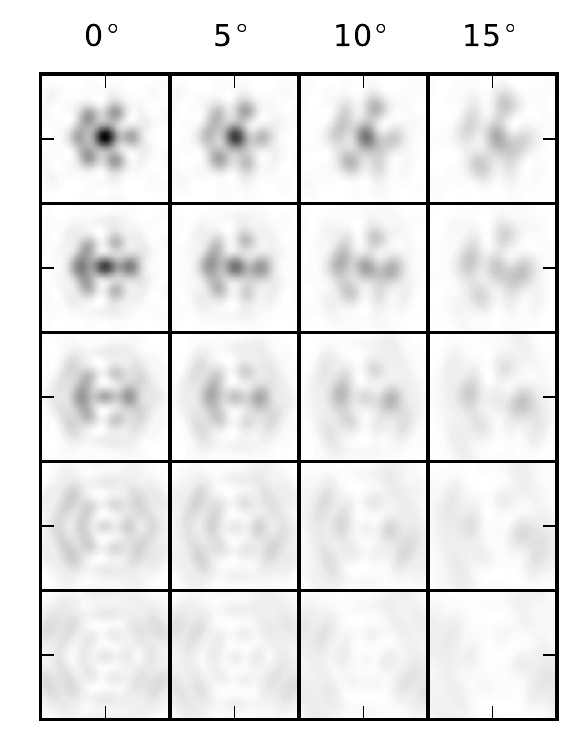}\hfill{}
\includegraphics[scale=0.85]{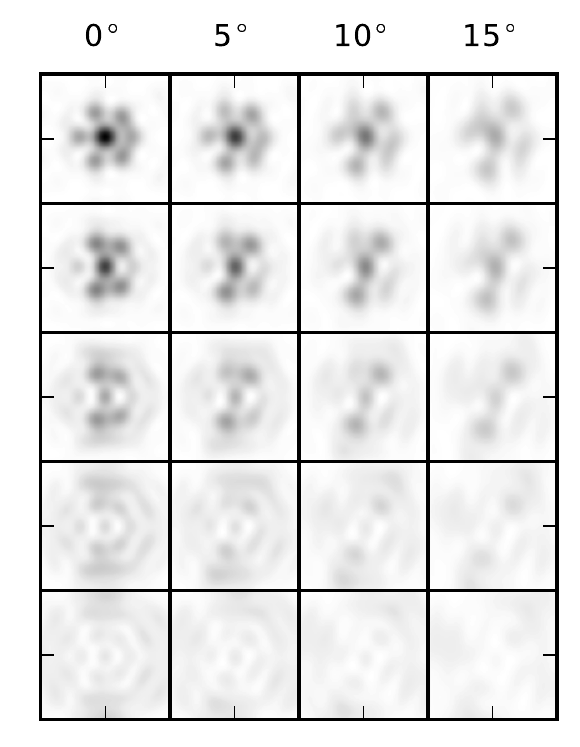}
\caption{FFDPs produced by a TIR CCR with a radial thermal gradient. Panels follow the conventions of Figure~\ref{fig:ffdp-axial-grid} with the exception of the center-to-edge temperature difference increasing in steps of 0.5~K.}
\label{fig:ffdp-radial-grid}
\end{figure}

In order to visualize the evolution of the FFDPs at various angles of incidence, we present $4 \times 5$ grids in Figures~\ref{fig:ffdp-axial-grid} and \ref{fig:ffdp-radial-grid} that vary the incidence angle in steps of 5$^{\circ}$ from left to right. Along the vertical axis, we show how the FFDPs evolve with incrementally increasing temperature differences. All irradiance values are normalized to the central irradiance of the normal incidence isothermal frame in the upper left. 

Figure~\ref{fig:ffdp-axial-grid} shows FFDPs for right-handed circular, horizontal linear and vertical linear input polarizations for the axial gradient case with the temperature difference increasing in 1~K increments. The patterns for left-handed circular input polarization are the same as that of right-handed except for a 180$^{\circ}$ rotation of each frame. Figure~\ref{fig:ffdp-radial-grid} depicts the FFDPs for the radial gradient case with the temperature difference increasing in 0.5~K increments. We see immediately that the central irradiance is quickly diminished when even modest thermal gradients are present, exacerbated by off-normal viewing.

\begin{figure}
\begin{center}
\includegraphics[scale=0.5]{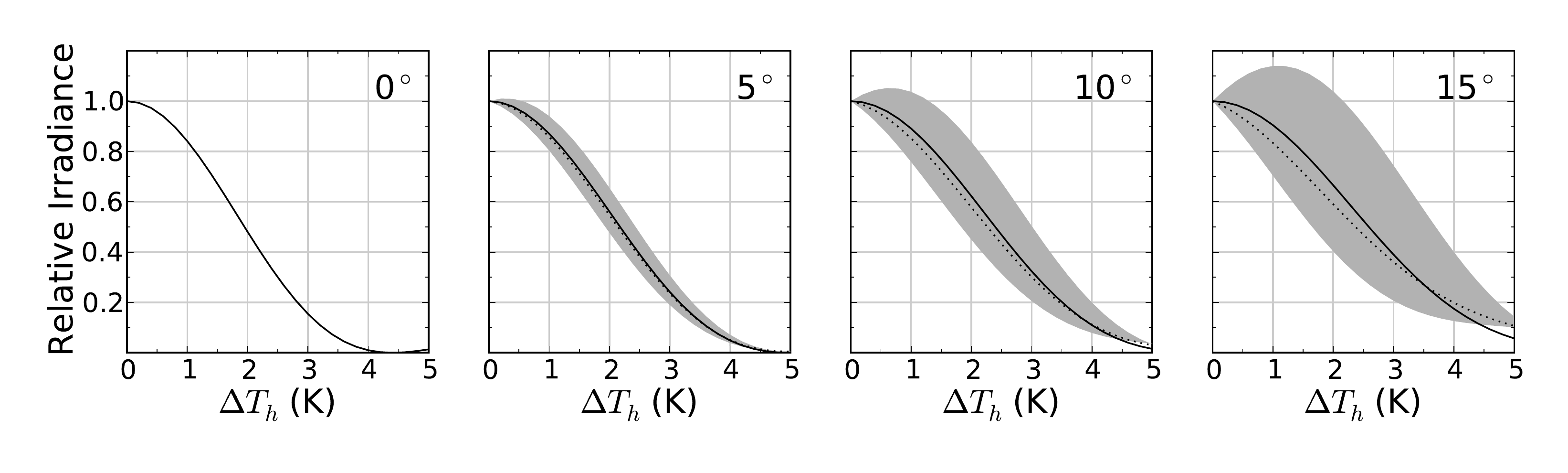}\hfill{}\end{center}
\caption{Central irradiance of the FFDP as a function of axial temperature difference. The solid line corresponds to a CCR employing perfectly reflecting (metalized) rear surfaces. For TIR, all rotations of linear input polarization fall within the gray envelope, and the dotted line corresponds to circular input polarization of either handedness. Incidence angle increases in $5^{\circ}$ increments from left to right and irradiance is normalized to the isothermal case in each panel. At normal incidence, the curves do not deviate from each other.}
\label{fig:central-axial-plot}
\end{figure}

\begin{figure}
\begin{center}
\includegraphics[scale=0.5]{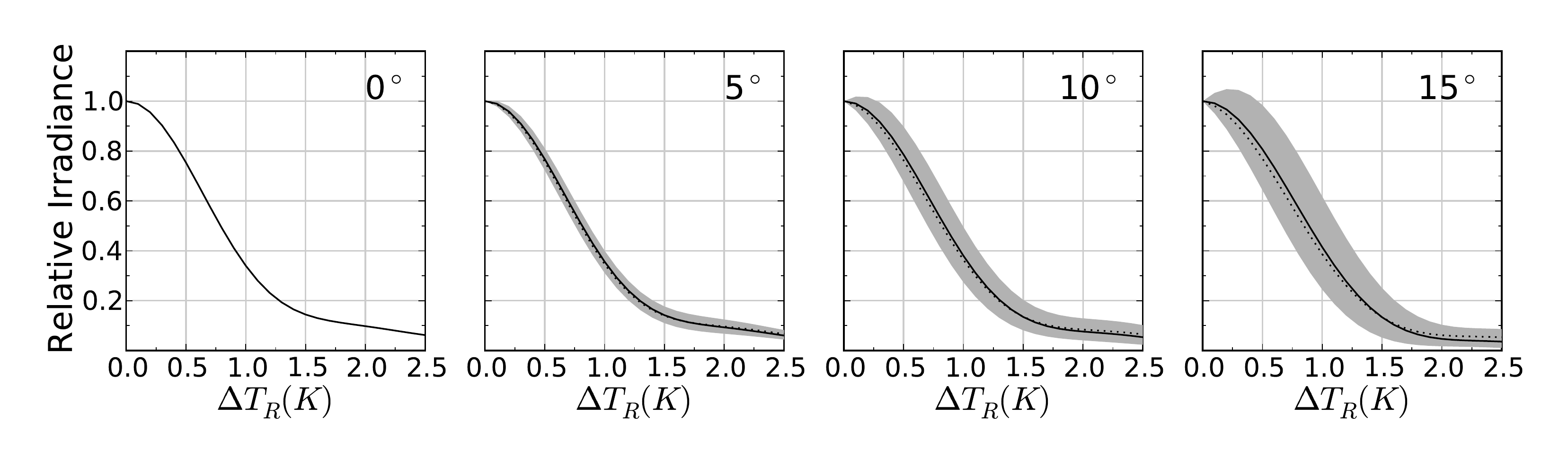}\hfill{}\end{center}
\caption{Central irradiance of the FFDP as a function of radial temperature difference. Panels follow the conventions of Figure~\ref{fig:central-axial-plot}. At normal incidence, the curves do not deviate from each other.}
\label{fig:central-radial-plot}
\end{figure}

In Figures~\ref{fig:central-axial-plot} and \ref{fig:central-radial-plot} we present graphs of the central irradiance as a function of temperature difference for key incidence angles. Again, irradiance is normalized to the isothermal case. The solid line in each panel corresponds to the response of a perfectly reflecting (metal-coated) CCR for arbitrary input polarization. For TIR, both types of circular input polarization appear as the same dotted line, and a gray envelope encompasses the range within which all linear input polarizations fall. We find that for isothermal TIR CCRs, the central irradiance is 26\% that of the perfect reflector case at normal incidence, and that this proportion holds precisely as thermal gradients are introduced. This demonstrates that the temperature dependence of the central irradiance is independent of polarization state at normal incidence.

Furthermore, at normal incidence, comparable radial and axial thermal gradients of 0.079~K~mm$^{-1}$ and 0.100~K~mm$^{-1}$ reduce the central irradiance to 15\% that of the isothermal case, respectively. In the radial gradient case, this is the start of the slowly-varying nearly-linear feature that begins at a temperature difference of 1.5~K.

While differing materials and temperature regimes will have different thermal coefficients, this will only cause a linear scaling of our results due to the linear dependence on $\eta$ in Equation~\ref{eq:phase-distortion-final}. For example, we see in Figure~\ref{fig:central-axial-plot} that the central irradiance exhibits complete destructive interference when the axial temperature difference is 4.4~K, corresponding to the spherical wavefront spanning exactly two Fresnel zones. If conditions double the thermal coefficient to $\eta = 2 \times 10^{-5}$~K$^{-1}$, the extinction occurs instead when the axial temperature difference is 2.2~K.  Likewise, the phase difference is inversely proportional to wavelength (Eq.~\ref{eq:phase-dif}).  A nulling of the central irradiance at $\Delta T_h\approx 4.4$~K at $\lambda_0=532$~nm would be pushed out to 5.2~K at a wavelength of 633~nm.

\subsection{Combined Axial and Radial Gradients}
\begin{figure}
\begin{center}
\includegraphics[scale=0.5]{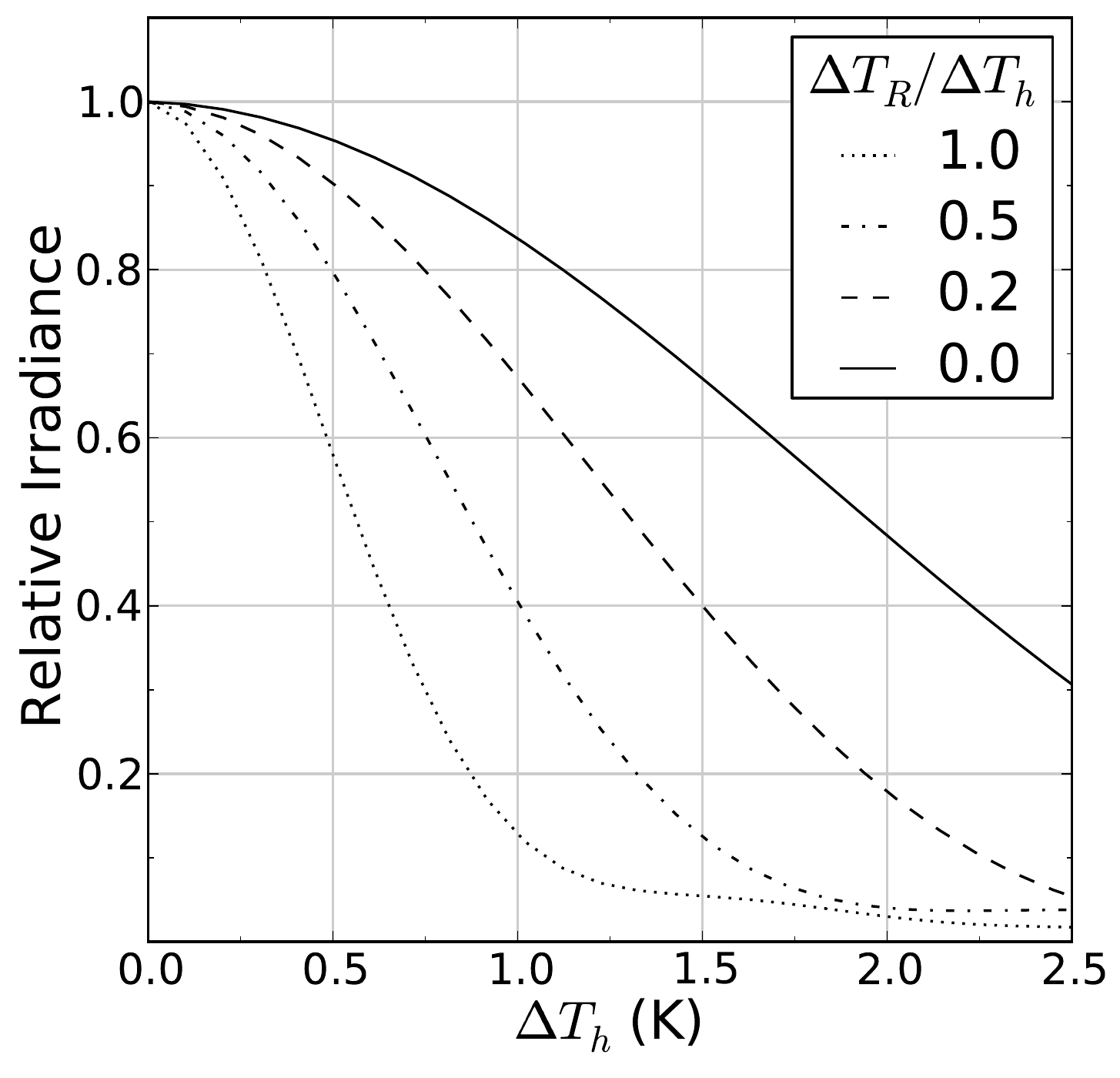}\end{center}
\caption{Central irradiance of the FFDP of a TIR CCR subject to both axial and radial thermal gradients, at normal incidence. As the axial temperature difference increases, the radial temperature difference follows proportionally. Irradiance is normalized to the isothermal case.}
\label{fig:plot-families}
\end{figure}

In Figure~\ref{fig:plot-families} we present smooth curves of the central irradiance in the presence of both axial and radial thermal gradients at normal incidence. As the axial temperature difference increases, we hold the ratio $\Delta T_{R}/\Delta T_{h}$ constant. For example, the dash-dot line represents the case where the radial temperature difference is half that of the axial temperature difference. Together, the two gradients work to reduce the central irradiance at smaller temperature differences. For reference, the solid line corresponds to a purely axial thermal gradient.

\subsection{Off-Center Irradiance}
\begin{figure}
\includegraphics[scale=0.4]{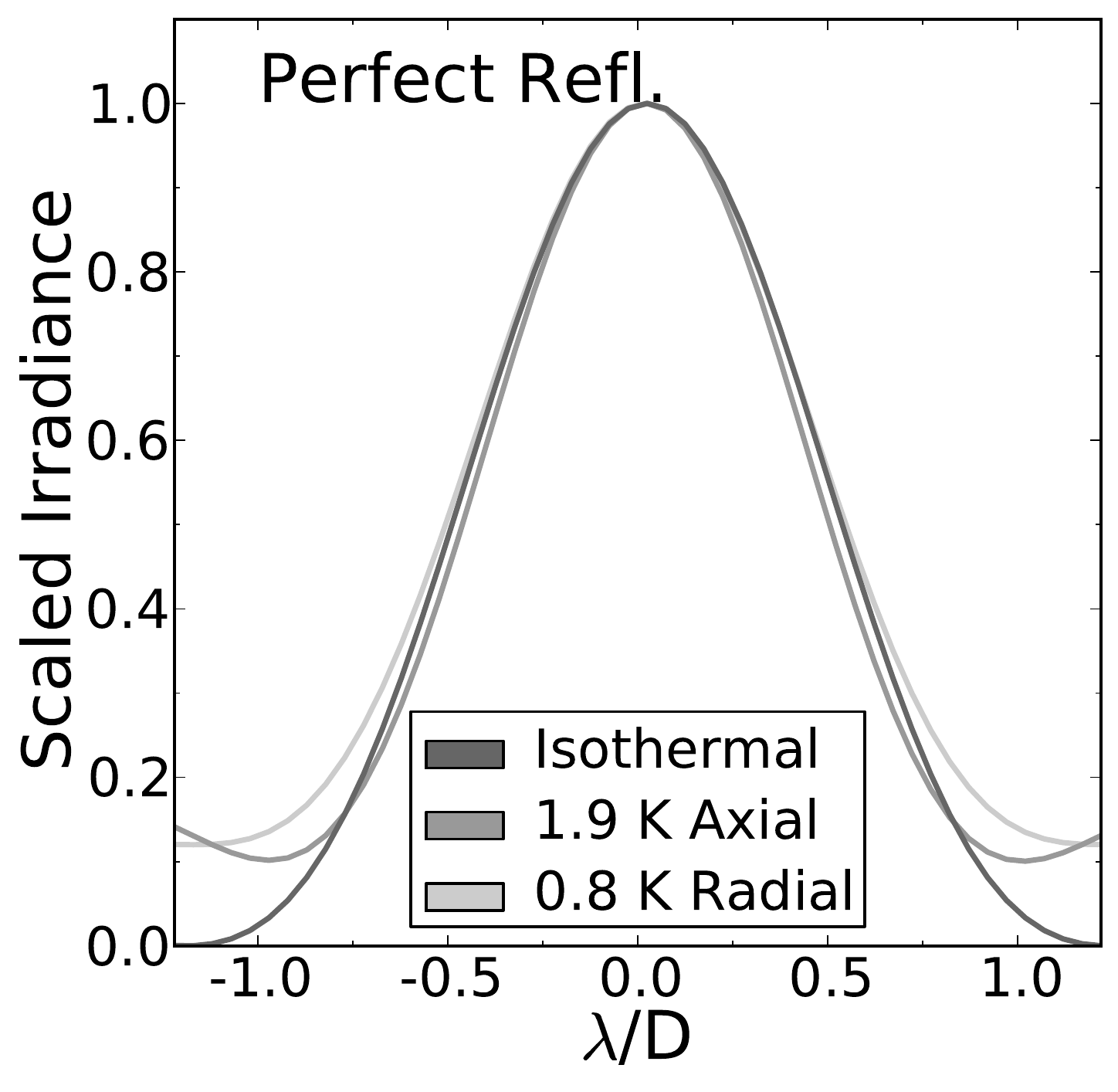}\hfill{}
\includegraphics[scale=0.4]{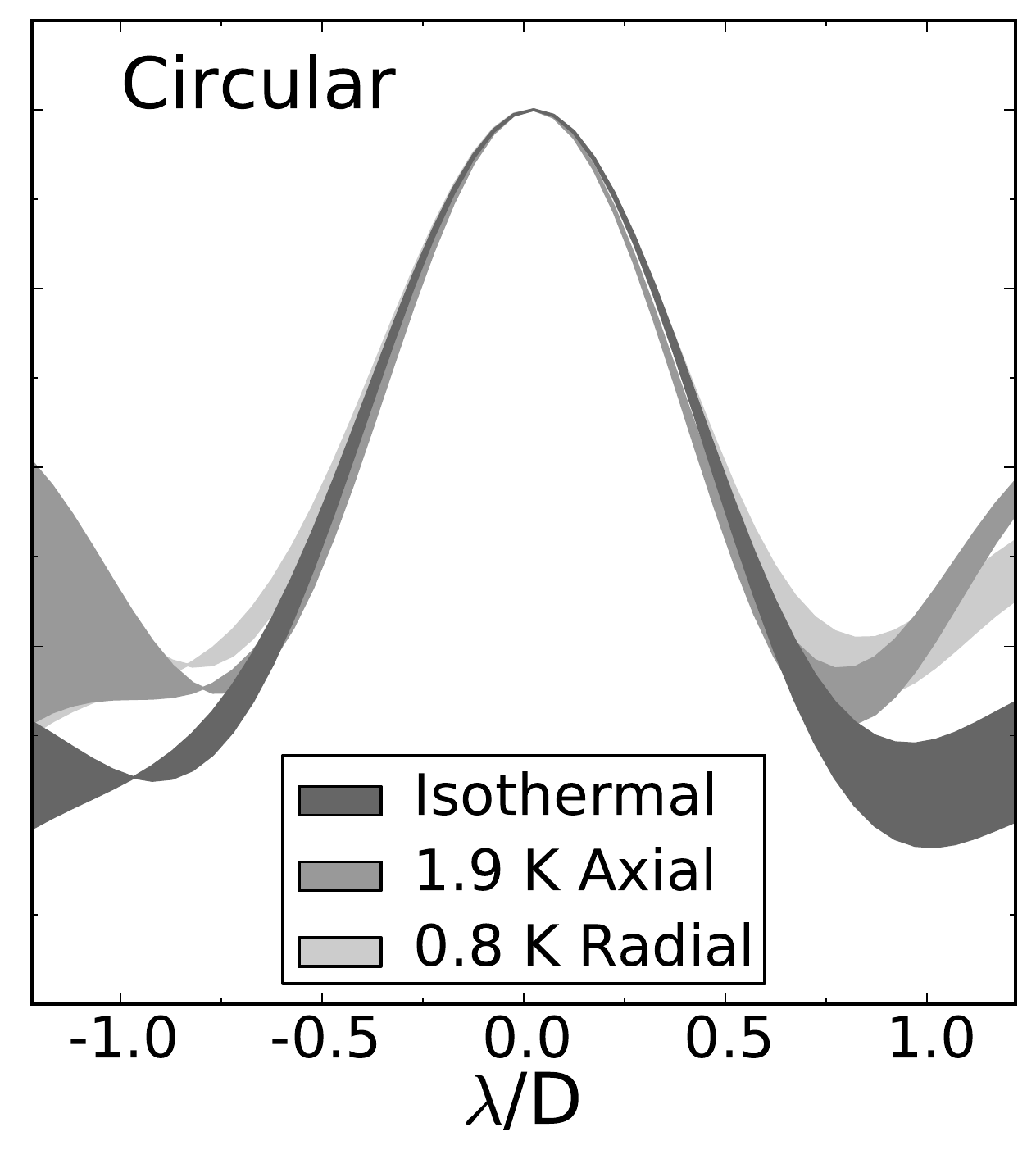}\hfill{}
\includegraphics[scale=0.4]{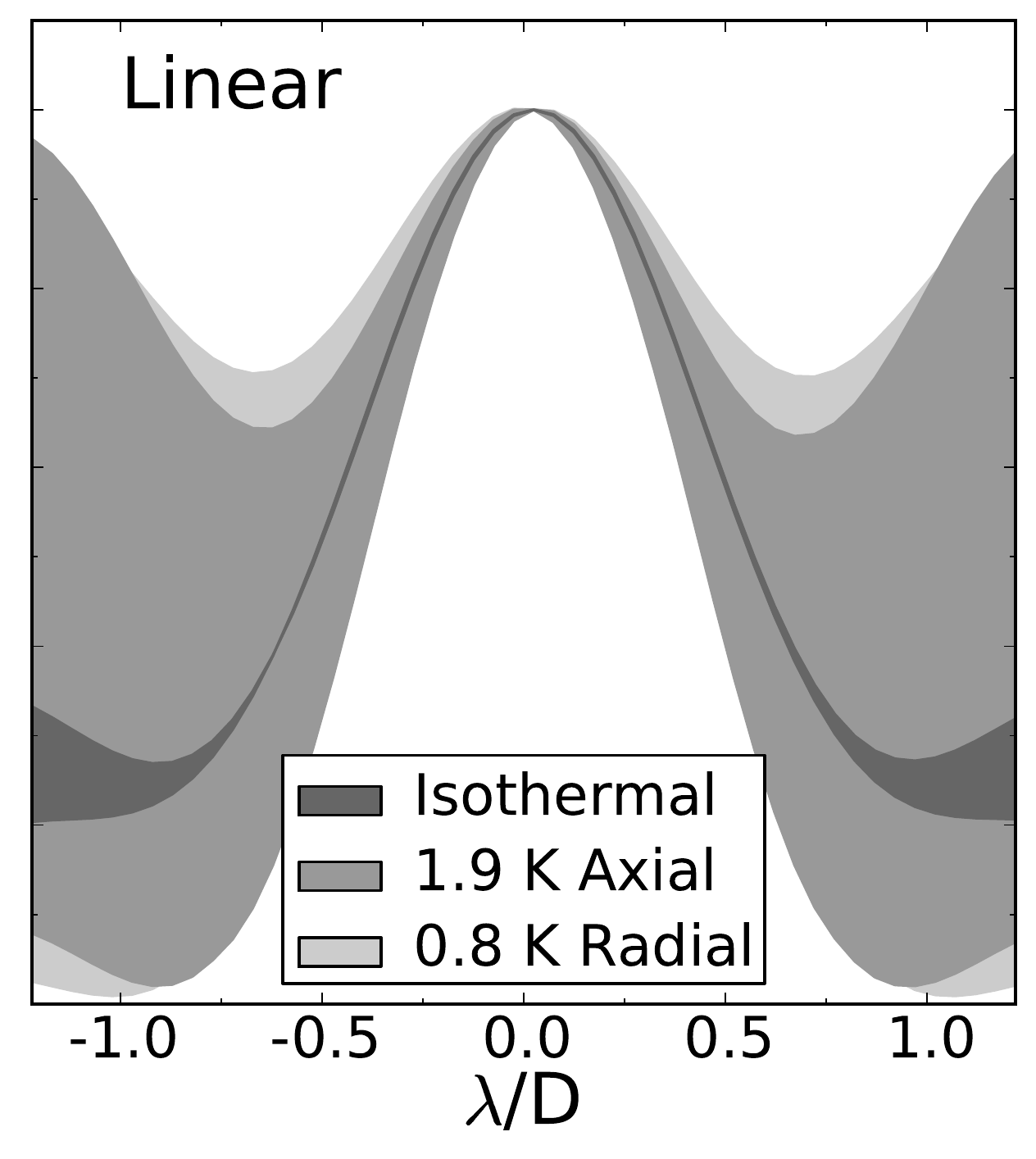}
\caption{Irradiance profiles of the central lobe of the FFDPs of TIR CCRs subject to thermal gradients at normal incidence. Temperature differences correspond to those required to reduce the central irradiance to half that of the isothermal case. The perfect reflector (metal-coated CCR) is at left, followed by TIR cases given right-handed circular and linear input polarizations. Envelopes encompass maximum variability in horizontal and vertical profiles, and for all rotations of the input polarization in the linear case. The isothermal case in the left panel corresponds to the Airy function.}
\label{fig:ref-slices}
\end{figure}

A prominent central lobe can be seen in the normal incidence cases of Figures~\ref{fig:ffdp-axial-grid} and \ref{fig:ffdp-radial-grid}. The shape of this lobe closely follows the Airy function in the isothermal case, as explored in our companion work \cite{raytrace}. In Figure~\ref{fig:ref-slices}, we show how the shape of the central feature evolves as a result of thermal gradients by plotting envelopes encompassing the maximum variability of horizontal and vertical profiles of the central region. We have selected temperature differences, $\Delta T_{h} = 1.9$~K and $\Delta T_{R} = 0.8$~K, which individually drive the central irradiance to half that of the isothermal case. Included is the perfect reflector (giving rise to the Airy function in the isothermal case), right-handed circular input polarization and linear input polarization. In the linear case, the envelopes include variation due to the angle of the input polarization as well. The profile of left-handed circular polarization is the same as that of right-handed except for a left-right flip.

From this we learn that all normal incidence cases mimic the Airy function near the center of the FFDP, to varying degrees, and that thermal gradients do not cause immediate departure from this similarity.

\section{Analytic Model of Central Irradiance}

Having seen how similar the central region is to that of the Airy function,
we can confine our interest to the central irradiance, knowing that we can
extend this result to points near the center of the diffraction pattern for
modest temperature gradients.  By doing so, we can analytically produce
useful approximations for the normal incidence case.

As outlined in our companion work \cite{raytrace}, the FFDP can be calculated from the Fourier transform of the complex amplitude and phase of the electric field exiting the CCR, with the square magnitude representing the irradiance in the far-field. A simpler expression may be employed to investigate the central value of the FFDP if one neglects off-axis field points:

\begin{equation}
I(0,0) = \left|\int\int_{\mathrm{aperture}} S(r,\theta) \exp \left[i\phi(r,\theta) \right] r \mathrm{d}r\mathrm{d}\theta \right|^{2}
\label{eq:ft-general-center}
\end{equation}

For TIR CCRs subject to thermal gradients, the phase function $\phi(r,\theta)$ can be represented as the superposition of the set of TIR static phase shifts, $\left\{\phi_{n}\left|\right.n:1\rightarrow6\right\}$, and the thermal lensing perturbations, $\phi_{\Delta T}(r,\theta)$, so that $\phi(r,\theta) = \phi_{n} + \phi_{\Delta T}(r,\theta)$---as depicted in the rightmost panel of Figure~\ref{fig:3d-wavefronts}. Additionally, as thermal gradients do not affect the amplitude of the electric field, the amplitude function is described entirely by the effects of TIR, making $S(r,\theta) = S_{n}$. Thus, 

\begin{equation}
I(0,0) = \left|
\int_{0}^{2\pi}\mathrm{d}\theta
\int_{0}^{R}r\mathrm{d}r S_{n} \exp \left(i\phi_{n}\right) \exp \left[i\phi_{\Delta T}(r,\theta)\right]\right|^{2}.
\end{equation}

Since $\phi_{n}$ and $S_{n}$ are constant within each wedge, we can pull them out of the integrals and sum over the six wedges:

\begin{equation}
I(0,0) = \left|
\sum_{n=1}^{6}\left\{ S_{n}\exp \left(i\phi_{n}\right)
\int_{\frac{\pi}{3}\left(n-1\right)}^{\frac{\pi}{3}n}\mathrm{d}\theta
\int_{0}^{R}r\mathrm{d}r \exp \left[i\phi_{\Delta T}(r,\theta)\right]\right\}\right|^{2}
\end{equation}

In Figure~\ref{fig:3d-wavefronts}, we saw that the thermal lensing perturbations are symmetric across each of the six wedges, allowing us to separate the integrals from the summation and integrate only over the first wedge, leaving

\begin{equation}
I(0,0) = \left|\sum_{n=1}^{6}\Biggl\{ S_{n}\exp \left(i\phi_{n}\right) \Biggr\}
\int_{0}^{\frac{\pi}{3}}\mathrm{d}\theta
\int_{0}^{R}r\mathrm{d}r \exp \left[ i \phi_{\Delta T}(r,\theta) \right]\right|^{2}.
\label{eq:separated}
\end{equation}

In the absence of thermal gradients, $\phi_{\Delta T}(r,\theta) = 0$ and the integral yields $\pi R^{2}/6$. For a perfectly reflecting CCR, the summation contributes a factor of 6 as $\phi_{n}$ does not vary and $S_{n} = 1$, therefore $I(0,0) = \pi^{2} R^{4}$. The phases, $\phi_{n}$, introduced by TIR, and detailed in \cite{raytrace}, result in $I(0,0) = 0.264\,\pi^{2} R^{4}$ for a fused silica corner cube, making the central irradiance 26.4\% that of the Airy function. Note that the integrals and summation are now separable, so that the normal incidence TIR response for nonzero thermal phase functions (displaying 6-fold symmetry) is

\begin{equation}
I_{\mathrm{TIR}}(0,0) = 9.505\left|
\int_{0}^{\pi/3}\mathrm{d}\theta
\int_{0}^{R}r\mathrm{d}r \exp \left[i \phi_{\Delta T}(r,\theta)\right]\right|^{2}.
\label{eq:ft-TIR}
\end{equation}

For the axial gradient case, we approximate the spherical wavefront as a paraboloid centered at the origin: $\phi_{\Delta T}(r) = k r^{2}$. We constrain this paraboloid to have a phase offset of $\phi_{R} = k R^{2}$ at the perimeter---which we will later relate to the thermal gradient---and find that 

\begin{equation}
I_{\mathrm{TIR}}(0,0) = 0.528 \, \pi^2 R^{4} \frac{1-\cos \left( \phi_{R} \right)}{\phi_{R}^2}.
\end{equation}

Similarly, we approximate the radial gradient case's pointed wavefront as a cone, $\phi_{\Delta T} = k r$. Applying a similar constraint, $\phi_{R} = k R$, we get:

\begin{equation}
I_{\mathrm{TIR}}(0,0) = 1.056 \, \pi^{2} R^{4} \frac{\phi_{R}^{2}-2\cos\left(\phi_{R}\right)-2\sin\left(\phi_{R}\right)+2}{\phi_{R}^{4}}
\end{equation}

\begin{figure}
\includegraphics[scale=0.4]{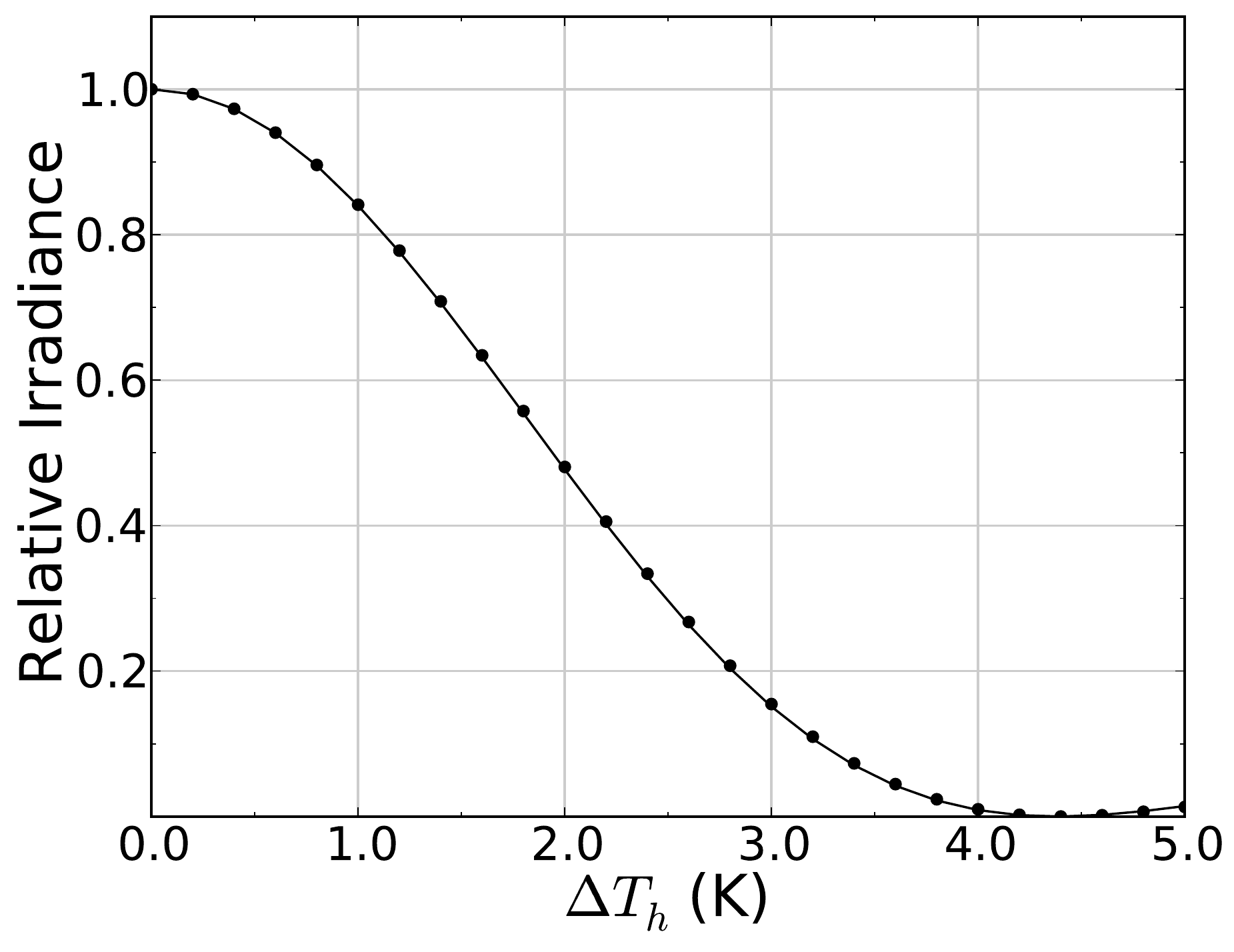}\hfill{}
\includegraphics[scale=0.4]{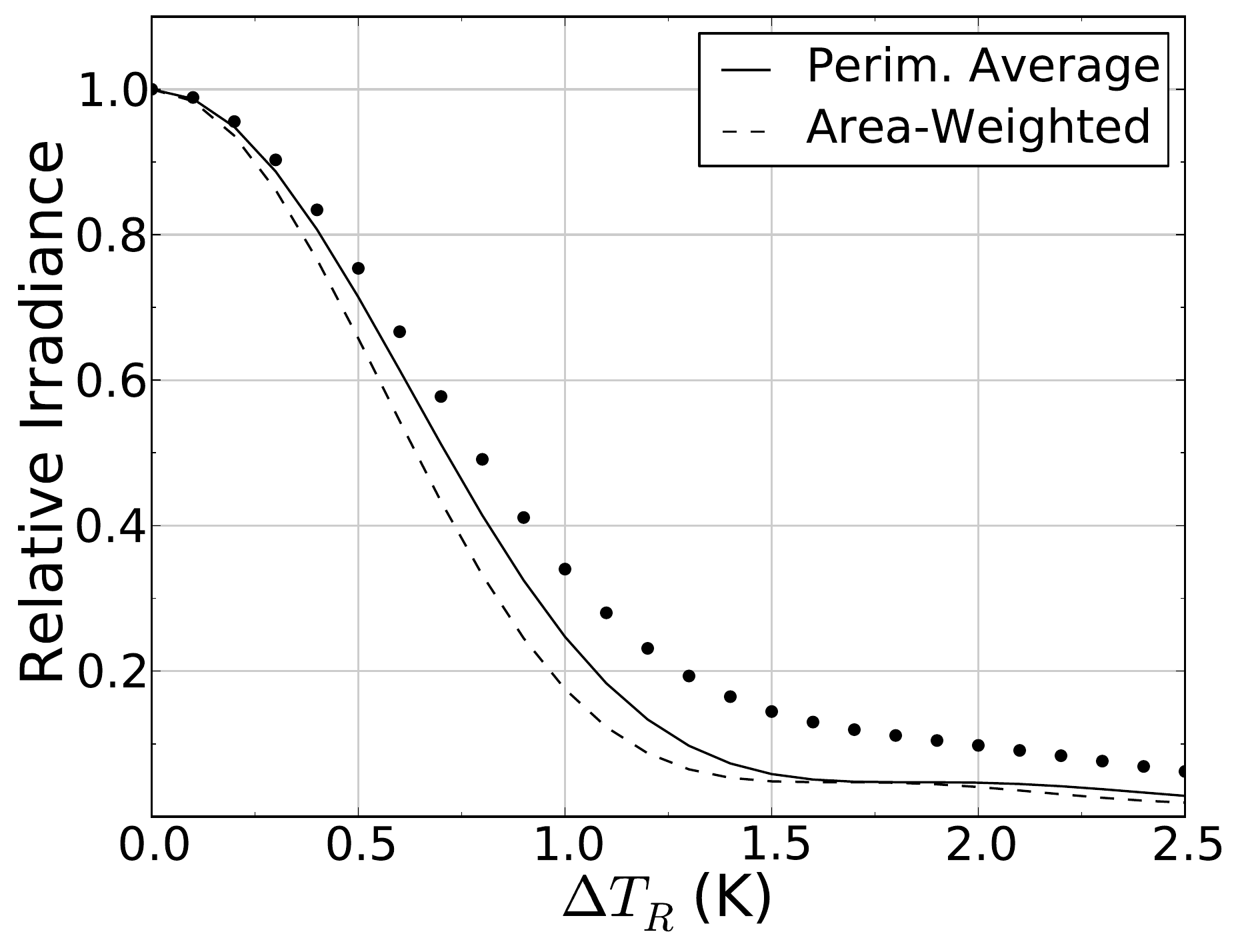}
\caption{Comparison of the central irradiance of simulated FFDPs (points) and the prescribed analytic models (lines), with axial thermal gradients at left and radial thermal gradients at right, for normal incidence. The axial wavefront is modeled as a paraboloid and the radial wavefront is modeled as a cone. Irradiance is normalized to the isothermal case.}
\label{fig:central-compare}
\end{figure}

A comparison of these models with our simulation results is shown in Figure~\ref{fig:central-compare}, with the axial thermal gradient model at left and the radial thermal gradient model at right. The axial gradient result is in strong agreement with that reported by Emslie and Strong \cite{nasa}, which has the same functional form and extinction point when appropriately scaled for the difference in wavelength.

Furthermore, we can combine Equations~\ref{eq:thermal-model} and \ref{eq:phase-dif} to determine the phase offset $\phi_{R}$ at the perimeter of the perturbed wavefront. The result simplifies greatly when we confine our interest to CCRs without a rim pad. In the axial gradient case, this phase offset corresponds to the uniform perimeter of the paraboloid in Figure~\ref{fig:3d-wavefronts}:

\begin{equation}
\phi_{R} = \sqrt{2} \, \frac{\pi \eta \Delta T_{h} R}{\lambda_{0}}.
\label{eq:axial-phase-model}
\end{equation}

\begin{figure}
\begin{center}
\includegraphics[scale=0.5]{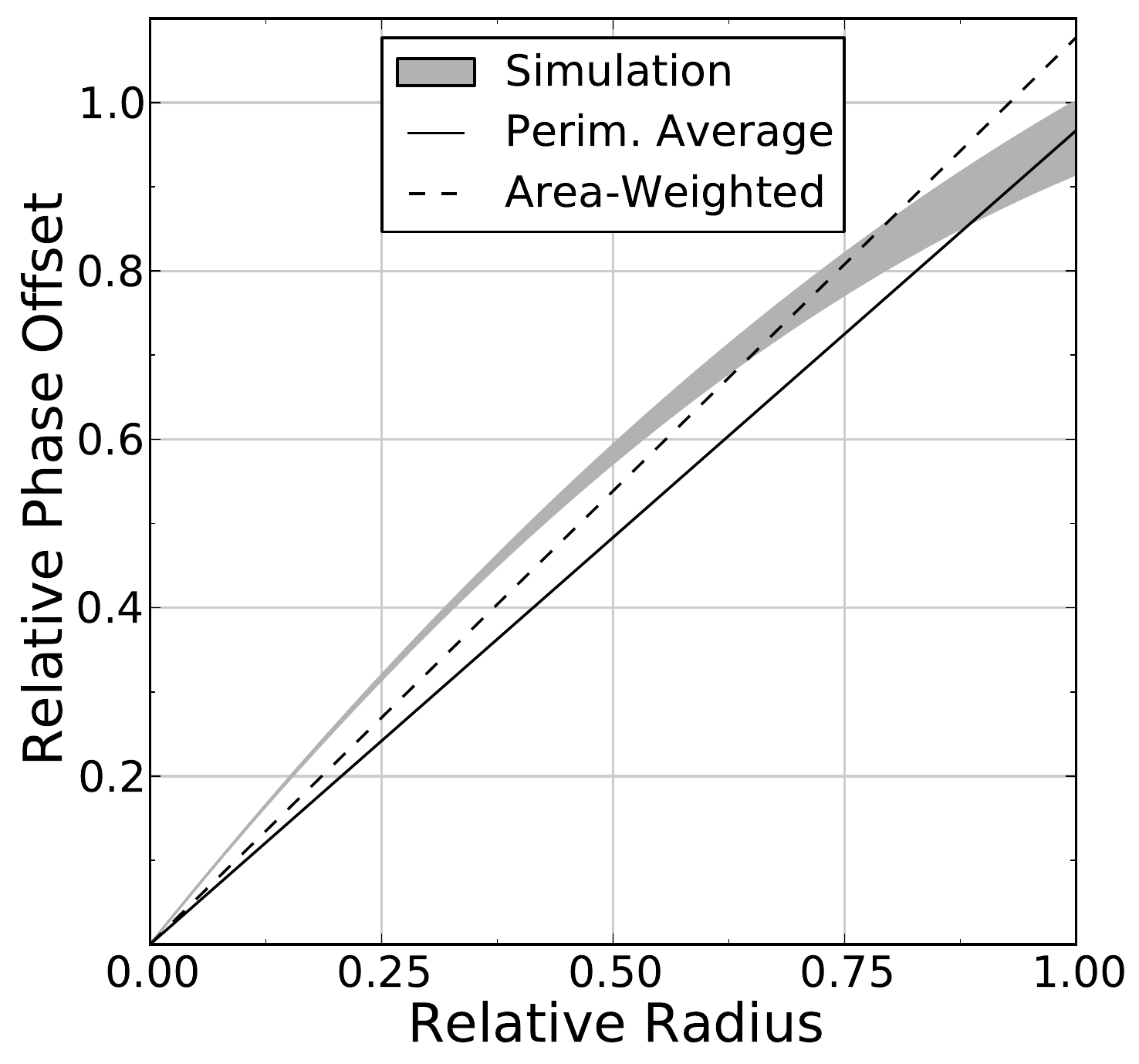}
\caption{Profiles of the radial thermal gradient phase perturbations and models. The gray envelope encompasses the variability of the wavefront perturbations. The solid line corresponds to the cone whose phase offset is the average of the extrema at the perimeter and the dashed line represents the cone which equally over- and under-estimates the area-weighted wavefront.}
\label{fig:radial-approx}
\end{center}
\end{figure}

The nonuniformity of the perimeter in the radial gradient case obscures an obvious scale for the cone. This variability is shown in Figure~\ref{fig:radial-approx} as a gray envelope. Here, we explore two cases. In the first case, we use the average of the extrema along the perimeter as the edge phase offset (solid). For the second case, we choose a scale that forms a cone with an equal amount of over- and under-estimation of the actual area-weighted wavefront (dashed). Both models are represented in Figure~\ref{fig:central-compare}.  The phase offset of the cone that extends to the average of the extrema is then:

\begin{equation}
\phi_{R} = 3.737 \, \frac{\pi \eta \Delta T_{R} R}{\lambda_{0}}.
\label{eq:radial-phase-model}
\end{equation}

For CCRs with a rim pad, the scale of the perturbed wavefronts changes slightly. If we consider the rim pad size as a fraction of the radius of the CCR, in the axial gradient case, the rim pad depreciates the value of $\phi_{R}$ by roughly 7\% for every tenth of the radius. In the radial gradient case, the rim pad appreciates the scale of the rim pad by about 12\% for every tenth of the radius. For example, the Apollo CCRs we are modeling have a rim pad of 2.9~mm, or 15\% of their radius. Here, we see a depreciation of the axial wavefront perturbation by 10\% and an appreciation of the radial wavefront perturbation by 17\%.  Applying these corrections brings the results of Eqs.~\ref{eq:axial-phase-model} and \ref{eq:radial-phase-model} in line with the numbers reported in Section~\ref{sec:method} for the Apollo corner cubes.

\section{Conclusions}

The introduction of axial or radial thermal gradients within CCRs amounting
to only a few degrees diminishes the central irradiance of the FFDP by an
order-of-magnitude.  The flux in the Airy-like central lobe of the FFDP is
well-represented by the central irradiance.  This result is qualitatively
independent of input polarization state, angle of incidence, or reflector
type.  Together,
minor thermal gradients along both principle axes drive the central
irradiance to near-extinction regardless of input polarization and angle of
incidence.

We provide generalized analytic models that approximate the behavior of
the central irradiance subject to axial or radial thermal gradients at
normal incidence, enabling easy determination of the sensitivity of
TIR CCRs in differing conditions. These models adequately capture the 
simulation results, as well as early models of the Apollo retroreflector
arrays, and provide quantitative guidance on tolerable thermal gradients 
in next-generation retroreflectors.  The Python code used to generate these results may be found at \url{http://physics.ucsd.edu/~tmurphy/papers/ccr-sim/ccr-sim.html}.

\section*{Acknowledgments}
Part of this work was funded by the NASA Lunar Science Institute as part of the LUNAR consortium (NNA09DB30A), and part by the National Science Foundation (Grant PHY-0602507).

\end{document}